\documentclass[11pt]{article}
\usepackage{latexsym}
\usepackage{epsfig}
\usepackage{amsmath}
\usepackage{graphics}
\usepackage{amssymb}
\usepackage{amsthm}
\usepackage{amsopn}
\usepackage{amscd}
\usepackage{fullpage}

\usepackage{setspace}
\newtheorem{thm}{Theorem}

\newtheorem{lem}[thm]{Lemma}

\newtheorem{defn}[thm]{Definition}

\numberwithin{equation}{section} 
\numberwithin{figure}{section}
\numberwithin{thm}{section}

\begin{document}
\begin{center}
\textbf{Finding Traitors in Secure Networks Using Byzantine Agreements}\\
\textbf{Liam Wagner\footnote[1]{Email: LDW@maths.uq.edu.au\\ Mailing Address: Department of Mathematics, The
University of Queensland, St Lucia, Brisbane Queensland 4072 Australia}\\Department of Mathematics and\\
ARC Centre for Complex Systems within\\ The University of Queensland}
\\
and
\\
\textbf{Stuart McDonald\footnote[2]{Email: s.mcdonald@mailbox.uq.edu.au}\\School of Economics and\\
ARC Centre for Complex Systems within\\The University of Queensland}
\end{center}

\pagenumbering{arabic}
\begin{abstract}
\noindent Secure networks rely upon players to maintain security and reliability. However not every player
can be assumed to have total loyalty and one must use methods to uncover traitors in such networks. We use the
original concept of the Byzantine Generals Problem by Lamport \cite{lam1}, and the more formal Byzantine Agreement
describe by Linial \cite{lin}, to find traitors in secure networks. By applying general fault-tolerance 
methods to develop a more formal design of secure networks we are able to uncover traitors amongst a group of players. 
We also propose methods to integrate this system with insecure
channels. This new resiliency can be applied to broadcast and peer-to-peer secure communication systems where
agents may be traitors or become unreliable due to faults. 

\end{abstract}


\section{Introduction}
A reliable communications system must be able to cope with failure of one or 
more of its components. Users within a communications network can also be classified 
as components of this system. A failed component may 
exhibit many different types of behaviour, which may include, sending 
conflicting, spurious or clearly false information. This sort of problem 
was expressed abstractly by Lamport \cite{lam1}, as the Byzantine Generals 
Problem (BGP).
\\\\
The best way to conceptualise the BGP is to use the example of an army poised for 
attack \cite{lam1}. The army is comprised of several divisions, each
commanded by a general. Having sent out observers the general must decide on a
course of action. This must be a collective decision based on all the available
facts and played out by each division in unison. However in some cases there
may be a traitor. 
\\\\
While broadcasting guarantees the recipient of a message that everyone else has
received the same message. This guarantee may no longer exist in a setting in which
communications are peer-to-peer and some of the people within this network are
traitors. In this type of setting a \textit{Byzantine agreement} offers a means
to achieving the required form of broadcast.
\\\\
Byzantine Agreements are used widely as a method for fault tolerance in
distributed systems. We have outlined the original literature
of Lamport \cite{lam1,lam2} and Pease \cite{pease} so that we can explore the
area in greater depth. The use of a more formalised version of the BGP was investigated in section 3,
using the developments of Pease and Lamport \cite{pease,lam1}. Furthermore we
also adapt a approximate solution to the infinite message case of the Weak
Byzantine Generals Problem of Lamport \cite{lam2}.
\\\\
We go on to develop the use of Byzantine Agreements (BA), in a
secure communication environment. Linial \cite{lin}, provides us with a wide
ranging insight into how BA's can be used to establish protocols for secure
communication. We also define current cryptographic methods in terms of a BA and
examine how these methods compare to information theoretic protocols.
\\\\
We move on to using BA's in an insecure environment, where communication channels can become faulty. Dasgupta's
\cite{das}, work on agreement using faulty interfaces, develops an analogy very
close to that of channels which may become unreliable.

\section{BA in Secure Communication}
The Byzantine Agreement problem is one of a collection of more general problems
in Fault-tolerance. In this section we apply the work of \cite{lin,ben,gold}, in an attempt to further our case for applying
Byzantine Agreements to secure communication and fault detection.

\subsection{Traitor-tolerance under secure communication}

\noindent Before we begin we need to make two assumptions about the behaviour of traitors. 
There are two types of bad player in this model, \emph{curious} or
\emph{malicious}. 

\begin{itemize}
\item Curious players try to extract as much information from the
fringes of operation as possible from exchanges from good players and
themselves. This raises the problem of information leaks, and trying to prevent
curious players from taking advantage of this source information. 
\item Malicious act in a manner which can directly undermine the integrity of a network.
\end{itemize}

\noindent There are two models for how good players hide information:
\begin{itemize}
\item \textbf{Information-theoretic:} Secure communication channels exist between every two agents. No third party can gain any information by eavesdropping messages sent on any such channel. 
A good example of this sort of protection against a man in the middle attack such as this, is the use of
quantum cryptography over fibre optic cables.  

\item \textbf{Bounded Computational Resources/Cryptographic set-up:} It is assumed in these models that the 
participants have restricted computational power. For example:

\begin{itemize}
\item Secure message passing: This case is only of interest over Insecure
channels and/or if the bounds on computational power allow the simulation of a 
secure communication channel.
\item The "and" function: Two players have a single input bit each, and they
need to compute the logical "and" of these two bits. Secure channels do not 
help in this problem, but this task can be performed in the cryptographic set-up.
\item The millionaires' problem: There is a protocol which allows players $P_{1}$ and $P_{2}$ to find out which of the integers $x_{1},x_{2}$ is
bigger, without $P_{1}$ finding out any other information about $x_{2}$ and vice versa? 
This is only interesting if there is a commonly known upper bound on both $x_{1},x_{2}$.
\item Game playing without a Grand Designer: Barany and F$\ddot{u}$redi \cite{bar}
show how $n\geq4$ players may safely play any noncooperative game in the absence of a
grand designer, even if one of the players is trying to cheat. As Linial outlines in \cite{lin},
this result can be strengthened, so that this condition will hold even if as many as $\lfloor \frac{(n-1)}{3}\rfloor$ 
players deviate from the rules of the game.
\item Secure Voting: Consider $n$ voters, each of whom casts one 
$yes/no$ vote on an issue. At the end of the voting 
round we may ask that the tally be made known to all players. This observation
should be taken into account in making the formal definition of 
"no information leaks are allowed". 
\end{itemize}
\end{itemize}
\noindent Based on Linial's work \cite{lin}, we investigate the two main models for bad players' behaviour.
\\\\
\textbf{Model 1: Curious Players} 
\begin{itemize}
\item Store all messages seen throughout the duration of the protocol.
\item Traitors collaborate to extract as much information as possible from their records of a run.
\item The behaviour of players who are said to be \textit{curious}.
\item We must impose a \emph{No Information Leak} clause on this model, so that no information other than that collected
by the traitors as a group is stored to undermine the network.
\end{itemize}

\noindent \textbf{Model 2: Malicious Players} A more demanding model assumes that nothing with regard to the behaviour of the 
traitors as in the Byzantine Agreement problem. In this situation we are more concerned with the \textit{correctness} of the computation is in jeopardy. If we were to compute $f(x_{1},\dots,x_{n})$ and some player $i$ refuses to reveal $x_{i}$ (which 
is known only to him), then any calculation dependant on this is corruptible. Furthermore if player $i$ intentionally sends an incorrect value for $x_{i}$, they would be doing so with the desire of being undetected. If it were possible to relax the requirement for no information leak, then correctness can in principle be achieved through the following \textit{commitment} mechanism:

\begin{itemize}
\item If each player places their value for $x_{i}$ in an envelope then all envelopes are publicly opened and referred to be locally threaded.
\item We can evaluate $|f(x_{1},\ldots ,x_{n})|- f(x_{i},\tilde{x}_{i})|<\varepsilon$   so that if $x_{i}$ is not valid we can draw from the set 
$z_{1},\dots,z_{n}$ for a valid response.
\end{itemize}
Thus we can perform these tasks without using physical envelopes. 
Given appropriate means for concealing information, as well as an 
upper bound on the number of faults, it is possible to compute both 
correctly and without leaking any information.
\\\\
The type of protocols that may be applied can be classed as follows:
\begin{itemize}
\item A specification of the task which is to be performed.
\item An upper bound $m$ for the number of unreliable players out of a total of $n$
\item The assumed nature of the traitors; \textit{curious} or \textit{malicious}
\item The countermeasures available: Either secure communication lines or a bound on the disloyal players' computational power.
\end{itemize}
The main result of this section, is that if the number $m< \infty$ of traitors
is properly bounded, so that both modes of deceit (curious and malicious) with
two guarantees of safety (secure lines and restricted computational power) enable 
for correct and leak free computation. We must now state how these results applies to our
problem from Linial \cite{lin} :
\\\\
\textbf{(1):} Given $f:f_{1},\dots,f_{n}$, in $n$ variables and players $P_{1},\dots,P_{n}$
which communicate via secure channels then each player $P_{i}$, holds an input $x_{i}$ known only to them. There exists a protocol
which is leak free against any proper minority of curious players. Given a coalition of players $S \subseteq \{1,\dots,n\}$ with $|S| \leq \lfloor \frac{(n-1)}{2} \rfloor$, where every communique is computationaly 
based on the set of messages passed to any $P_{j}(j\in S)$ can also be computed given only the $x_{j}$
and $f_{j}(x_{1},\dots,x_{n})$ for $j \in S$.
\\\\
The computation of $f(x_{1},\dots,x_{n})$ is not guaranteed to force
traitors to supply their correct input values. The best that can be 
hoped for is that traitors can be made to commit on input values, which are independent of the input reliable 
players. After such a commitment stage, the computation of $f$ proceeds correctly and without leaking information. 
In any case a traitors refusal to supply an input, will result in the default value. That is, the protocol computes a value $f(y_{1},\dots,y_{n})$ so that $y_{i} = x_{i}$ for all $i \notin S$ and where the $y_{j}(j \in S)$ are chosen independently of $x_{i}(i\notin S)$.
\\\\
The same results hold if the functions $f_{j}$ are replaced by probability distributions and instead of computing the functions we need to sample according to these distributions. We should also restate that the bounds on this theorem are indeed tight.
\\\\
\textbf{(2):} Assume that one-way trapdoor permutations exist, p1356 \cite{lin}. If we modify the situation in (1) as follows we can use the following:
\\\\
Channels are not secure, but agents are probabilistic, polynomial-time Turing Machines. Similar conclusions hold with the bound $\lfloor \frac{(n-1)}{2} \rfloor$ and $\lfloor \frac{(n-1)}{3} \rfloor$, replaced by by $n-1$ and $\lfloor \frac{(n-1)}{2} \rfloor$ respectively. Again the bounds are tight and the results hold also for sampling distributions rather than for evaluation of functions \cite{lin}.

\subsection{Protocols for Secure Collective Communication}
Given a set of $n$ agents and an additional trusted party which may be referred to as a grand designer \cite{dod}, there are various goals that can be achieved in terms of correct, reliable and leak-free communication. In fact, 
all they need to do is relay their input values to the party who can compute any functions of these inputs and 
communicate to every player any data desired. In broad terms, our mission is to provide a protocol for the $n$ 
parties to achieve all the tasks in the absence of a trusted party. The two most important instances of this 
general plane are:
\\\\
\textbf{Privacy:} If we consider a protocol for computing $f_{1},\dots,f_{n}$, where originally party $i$ holds $x_{i}$, the value of the $ith$ variable and where by the protocol's end it is known that $f_{i}(x_{1},\dots,x_{n})(1\leq i \leq n)$. 
The protocol is \textit{t-private} if every quantity which is computable from the information viewed throughout the protocol's 
run by any coalition of $|S|$ players, is also computable from their own inputs and outputs.
\\\\
\textbf{Fault tolerance:} The protocol is \textit{t-resilient} if for every coalition $S$ of no more than $t$ parties such that
$|S| \leq t$ and the protocol computes a value $f(y_{1},\dots , y_{n})$ so that $y_{i}=x_{i}$ for all $i\notin S$ and so that 
$y_{j}(j\in S)$ are chosen independent on the value of $x_{i}(i\notin S)$.
\\\\
We shall now express the results which hold under the assumption that traitors are coeducationally restricted.
\begin{thm}
Every function of $n$ variables can be computed by $n$ agents which 
communicate via secure channels in a  $\lfloor \frac{(n-1)}{2} \rfloor$-private way. 
Similarly, a protocol exists which is both  $\lfloor \frac{(n-1)}{3} \rfloor-$private 
and $\lfloor \frac{(n-1)}{3} \rfloor-$resilient, where the computational bounds are tight.
\end{thm}
\noindent The are four protocols to describe according to the model (cryptographic or information-theoretic), where bad players are assumed curious or malicious. All four protocols follow one general pattern, which is explained below. We review the solution for the case in reasonable detail and then indicate how it is modified to deal with the other three solutions. 
\\\\
Since circuits can simulate Turing Machines, the problem becomes more structured, when rather than dealing with
a general function $f$ the discussion focuses on a circuit which computes it with-out loss of generality. In Goldreich \cite{gold}, the idea that players collectively follow the computation carried out by the circuit moving from one gate to the next, but where each of the partial values computed in the circuit, is encoded as a secret shared by the players. To implement this idea one needs to be able to:
\begin{itemize}
\item Assign input values to the variables in a shared way.
\item Perform the elementary field operations on values which are kept as shared secrets. The outcome should again be kept as a shared secret.
\item If, at the computation's conclusion, each player $P$ is to possess a certain value computed throughout, then all shares of this secret are to be handed to him by all other players.
\end{itemize}
In Linial \cite{lin}, the 2nd item is investigated at greater depth, and we
shall now follow his reasoning. If we reconsider the information-theoretic, curious player scenario, we need to carry out our
investigation in the following manner. Secrets are shared using a digital signature,
and we need to be able to add and multiply field elements which
are kept as secrets shared by all players.
\\\\
Schamir \cite{schamir}, goes on to describe the importance of dealing with the
degree being too high, which thus needs to be reduced. This is achieved by
truncating high-order terms in $g$ (g is the secret). Letting $h$ be the polynomial obtained by deleting all terms in $g$  of degree exceeding $m$ (m is the number of players). If \textbf{a} (and
respectively \textbf{b}) is the vector whose $ith$ coordinate is $g(\alpha_{i})$
[respectively $h(\alpha_{i})$] the there is a matrix $C$ depending only on the
$\alpha_{i}$ such that $\textbf{b}=\textbf{aC}$. Thus a degree reduction, may be
performed in a shared way, as follows. Each $P_{j}$ knows $g(\alpha_{j})$ and for
every $i$ we need to compute $h(\alpha_{i}) = \sum_{j} c_{i,j}g(\alpha_{j})$ and
inform $P_{i}$. Now, $P_{j}$ computes $c_{i,j}g(\alpha_{j})$ and deals it as a
shared secret among all players. Everyone then sums his shares for 
$c_{i,j}g(\alpha_{j})$ over all $j$, thus obtaining his share of
$h(\alpha_{i})=\sum c_{i,j}g(\alpha_{j})$, which becomes a shared secret. We
should recall from both Schamir and Linial \cite{schamir,lin}, that if $s_{v}$
are secrets, and $s^{\mu}_{v}$ is the share of $s_{v}$ held by player $P_{\mu}$
then his share of $\sum s_{v}$, is $\sum s^{\mu}_{v}$. So each player passes to
$P_{i}$ the share of $h(\alpha_{i})$, so now $P_{i}$ can reconstruct the actual
$h(\alpha_{i})$. Now free term of $g$ which is the same as the free term of $h$,
is kept as a shared secret, as needed. 
\\\\
This establishes the $\lfloor \frac{(n-1)}{2} \rfloor-$privacy part of the previous theorem.
The condition that $n > 2m$ is implicit, \cite{lin}. Linial \cite{lin} also
states that the more curious players cannot be tolerated by any protocol follows
from Chor's result that it is impossible to compute the logical "or" function
for two players \cite{chor}. Having dealt with \emph{curious} players, we shall
simply refer the reader to Linial's treatment of \emph{malicious} players,
(Linial \cite{lin}). 
\\\\
However we must state two important results from Linial which are directly
related to our secure communication theme for this paper. Besides the
corruption of shares, there is also a possibility that bad players who are
supposed to share information with others will fail to do so. This difficulty is
countered by using a verifiable secret sharing scheme \cite{men,schamir}.
\\\\
Secondly the \emph{malicious} case states that only $n\geq 3m+1$ can be 
dealt with in this way. This follows the Byzantine Agreement protocol
from section 3. So if players are not restricted to communicate via a secure
two-party line, but can also broadcast messages, fact can increase
resiliency from $\lfloor \frac{(n-1)}{3}\rfloor$ to $\lfloor \frac{(n-1)}{2} \rfloor$, \cite{lin}.

\section{BA and Insecure Communication Channels}
Insecure communications channels also provide challenges when trying to achieve
agreement amongst a group of players. We shall use the modified version of the
BGP to find agreement amongst a set of players who are reliable but may
encounter faulty interfaces \cite{das}. This is analogous to a faulty or insecure 
communication channels. In this section we will assume that a faulty channel 
can not be relied upon as being secure. We consider three types of faults, namely 
message corruption, message loss, and spurious message generation. A spurious message in our set of
circumstances would be regarded as a traitor generating some alternate set of messages to distribute across the
network.
\\\\
We shall use Dasgupta's \cite{das}, variant of the Byzantine
generals problem, where agents who represent traitors are fully operation.
However the disloyal agents interfaces with the communication network may be
faulty, causing them to send erroneous messages throughout the network. This model
can be briefly described by the following:

\begin{itemize}
\item Each agent has one or more communication devices available.
\item In order to send a message, the source agent passes the message to the
appropriate communications device.
\item A channel receives a message from the first agent and delivers it to
the second agent.
\item One or more devices may be faulty.
\item The agents themselves are reliable. However an agent with one or more faulty
devices is called the traitor.
\end{itemize}
We will use the conventions as outlined  in Dasgupta \cite{das} and categorize the
types of faults that are possible in this model as follows:
\begin{itemize}
\item  $m/m'$ fault: The device receives a message m and communicates a
different message $m'$ to the other agent.
\item $m/\theta$ fault: The device receives a message $m$ and loses the message.
\item $\theta/m'$ fault: The device generates a spurious message m and loses it.
\end{itemize}

\noindent These protocols are used to analyze the Byzantine Agreement in
the presence of faults. Throughout this report we will make the assumption
that is used in Dasgupta \cite{das} and in the were all inter-agent communication is
synchronous.
\\\\
So if we assume that all three types of faults are possible, then the agreement
problem reduces to the Byzantine Generals Problem such that more than two thirds
of the participants are required to be loyal. We should also note that if only
$m/m'$ faults are possible, then the agreement problem becomes trivial. In
Dasgupta \cite{das}, a protocol which achieves agreement in one round is presented.
Furthermore Dasgupta \cite{das} also shows that spurious messages causes the main
difficulty in reaching some sort of agreement. If $m/m'$ and $m/\phi$ are the
only possible faults, then Dasgupta \cite{das}, asserts that agreement is possible
irrespective of the number of traitors within the network. Using the protocol
outlined in Dasgupta \cite{das} we can use the proof of $n < 3m+1$ to argue that
these types of faults do not require interactive consistency.

\subsection{Agreement Under Faults}

\subsubsection{Agreement under $m/m'$, $m/\phi$ and $\phi/m$ faults}
In Dasgupta \cite{das} it can be seen that if $m/m'$, $m/\phi$ and $\phi/m$ faults are
all possible then the agreement is logically equivalent to the BGP. We will now
follow the work of Dasgupta \cite{das} and examine the ways in which an agent may fault
in the original BGP and demonstrate possible equivalent situations in this model:
\begin{enumerate}
\item A traitor receives a message and communicates some other message. This is
equivalent to the $m/m'$ fault.
\item A traitor receives a message and transmits nothing. ($m/\phi$ fault)
\item A traitor receives no message and transmits a spurious one. ($\phi/m'$
fault)
\end{enumerate}

Having observed that it is easy to see how agreements under $m/m'$, $m/\phi$
and $\phi/m$ faults is as difficult as the original BGP. We should also note that
the reverse is easier to see as the original model asserts that the agent could
be faulty. We shall now conclude this aspect of Dasgupta \cite{das}'s work by proposing a
short theorem with an easy proof.

\begin{thm}
If $m/m'$, $m/\phi$ and $\phi/m$ faults are all possible, then agreement is
possible in $m+1$ rounds amongst at least $3m+1$ agents.
\end{thm}

\textbf{Proof:} Follows from the equivalence with the original Byzantine
agreement problem. Agreement can be reached in $m+1$ rounds using the unforgeable
(oral) message protocol of Lamport in \cite{lam1}.

\subsection{The Importance of $m/\phi$ and $\phi/m'$ faults}

\textbf{Protocol for m/m'-only model}
\begin{enumerate}
\item The General decides whether to attack or retreat
\subitem If the decision is to retreat the general remains silent
\subitem If the decision is to attack, the general sends a message to all lieutenant
\item If a member of the network (other than the general) receives any message in the first round it decides to attack, otherwise it decides to retreat
\end{enumerate}

\begin{thm}
The protocol for the m/m' only model achieves agreement in one round.
\end{thm}

\noindent \textbf{Proof:} We shall outline a slightly different approach to that in
Dasgupta \cite{das}, and separate the two cases in a more formal way.

\begin{enumerate}
\item[I.] Suppose the commanding general decides to retreat. Since the general and his Lieutenant are correct, the general does not attempt to send any message. Since $\phi/m'$ problems are ruled out, none of the other members receive any message from the general and therefore all of the lieutenant retreat.
\item[II.]Now if we assume that the general decides to attack, then the entire network is correct. The general attempts to send a message to every other member of his forces. Since $m/\phi$ faults are ruled out, each of the Lieutenant receive some message (which may or may not be complete) from the general, and decide to attack.
\end{enumerate}

This result shows that if the Lieutenant's are themselves correct, then the main difficulty in achieving agreement is in the presence of $m/\phi$ and $\phi/m'$ faults. We must also observe that the absence of $\phi/m'$ and $m/\phi$ faults, the generals with faulty interfaces also reach the same consensus.

\subsection{Agreement under $m/m'$ and $m/\phi$ faults}
In this section we consider a communication system/scenario which does allow $\phi/m$.
We will consider this case out of special interest. Quite often network interfaces become faulty only if sensitized when an attempt is made to send messages through them. we present the algorithm as outlined in \cite{das}, which achieves agreement in at most $n+1$ rounds where n is the number of generals with faulty interfaces.
\\\\
In this protocol the decision to retreat is modelled by silence and attack is communicated by sending a single message to each participant. The protocol among n generals is recursively described by the following.
\\\\
\textbf{Algorithm} $M(0,n)$.
\begin{enumerate}
\item The Commanding General, communicates a message to every other general if it has decided to attack. Otherwise it remains silent.
\item Each of the generals, $G_{i}$, acts as follows. If $G_{i}$ has already decided, then it ignores all messages. If $G_{i}$ has not yet decided, then it decides to attack if it receives any message from the commander, and decides to retreat otherwise.
\end{enumerate}

\noindent \textbf{Algorithm} $M(k,n), k >0$
\begin{enumerate}
\item The commanding General communicates a message to every other general if it has decided to attack. Otherwise he remains quite.
\item Each of the other generals, $G_{i}$, act as follows. 
\begin{itemize}
\item If $G_{i}$ has already decided, then it will ignore all messages. 
\item If $G_{i}$ has not yet decided, then it decides to attack if it receives any message from the commanding general, and remains undecided otherwise. 
\item General $G_{i}$ now acts as the commander in algorithm $M(k-1,n-1)$ among the other $n-2$ generals.
\end{itemize}
\end{enumerate}

As we have seen in both Lamport's algorithm \cite{lam1} and in Dasgupta 
\cite{das}, the protocol starts when the general takes a decision on 
whether to attack or retreat, and initiates the protocol acting as the 
commander in algorithm $M(k,n)$. The following results establish that in
the presence of $m/m'$ and $m/\phi$ faults only, Algorithm $M(k,n)$ achieves
Byzantine Agreement in a cluster of $n$ agents, among these only $k$ agents may
have faulty interfaces.  Thus Byzantine agreement is possible in this model
irrespective of the number of agents that have a faulty interface.

\begin{lem}
If the instigator of the $M(k,n)$ algorithm decides to retreat, than all other
processors agree to retreat.
\end{lem}
\noindent \textbf{Proof:}  If the instigator decides to retest, then he sends no message in $M(k,n)$.
Since $\phi/m'$ faults are ruled out, as none of the agents receive any message,
and therefore send none in return. Thus in round $k+1$, when $M(0,n-k)$ is
initiated, so that all agents, including those with a faulty interface decide to
retreat.
\\\\
In this proposed algorithm of Dasgupta \cite{das}, all agents but for the
instigator, an agent sends out messages only if it receives a message in the
previous round. Thus except for the messages sent out by the instigator, each
message sent out by an agent is preceded (Causally) by the a receipt of a message by that
agent.
\begin{defn}
If an agent sends out a message $m'$ upon receiving a message $m$, then $m'$ is
referred to as being casually preceded by $m$. This relation is denoted $m\prec
m'$. Furthermore we can say that the casual precedence is transitive. Messages
which causally precede a message $m$ as being ancestors of $m$. We refer to the
set of agents constituting the sender of $m$ and the sender of all its ancestors
as the sender-set of $m$.
\end{defn}

\begin{lem}
If the first agent with all reliable interfaces reaches a decision to attack,
then the decision is made in round $j$, where $j \leq k$, then by the end of
round $j+1$, all agents with reliable interfaces agree to attack.
\end{lem}
\noindent \textbf{Proof:}
When an agent with reliable interfaces receives a message $m$, it decides to
attack, and in the following round it communicates messages to all the other
agents which is not a member of the sender-set of $m$. If $P$ is the first agent
with reliable interfaces to receive a message $m$ (and decides to attack). Thus
none of the agents in the sender-set of $m$ have all reliable interfaces.
Therefore, in the next round $P$ sends messages to all agents with reliable
interfaces, and each agent decides individually to attack.

\begin{lem}
If no agent with reliable interfaces reaches a decision to attack by round $k$,
then each agent with reliable interfaces will decide to retreat in round $k+1$.
\end{lem}

\noindent \textbf{Proof:}
Dasgupta \cite{das} shows that if none of the agents will all reliable
interfaces receive a message, and decide to attack by round k. Then none of the
agents receive a message in round $k+1$, therefore all of them decide to
retreat. The sender set of a message received in round $k+1$, has $k+1$ agents,
at least one of which must have reliable interfaces. That agent must receive a
message by round $k$. However this is a contradiction, since we have been given
the information that the agents with all reliable interfaces have received a
message by round $k$.

\begin{thm}
If $m/m'$ and $m/\phi$ are the only faults are possible, then it is possible to
reach Byzantine Agreement in a cluster of $n$ agents of which at most $k$ are
faulty/disloyal, irrespective of the ratio of $k$ and $n$. Therefore Agreement
can be reached in $k+1$ rounds.
\end{thm}

\noindent \textbf{Proof:}
We will use Dasgupta's proof, to show that the algorithm $M(k,n)$ achieves this
agreement. If $n-k \leq 1$, then the proof is self explanatory. Thus if we
consider the other case where the instigator decides to retreat. By Lemma 7.3,
all agents agree to retreat in round $k+1$. We shall now consider now the cases when
the instigator decides to attack. If we look at the two possible cases, which
depend on whether or not the interfaces of the instigator are all correct or
not. We shall follow Dasgupta \cite{das}, and treat each of these cases
separately.

\begin{enumerate}
\item [I.] The Instigator is reliable. Thus if all interfaces of the
instigator are reliable and the instigator decides to attack, then it is
successfully sends a message to all other agents in the first round.
\item [II.] The Instigator has faulty interfaces. If the instigator has one
ore more faulty interfaces and the instigator decides to attack, then it may
succeed in sending messages to some and none to others. In this case we need to
follow Dasgupta \cite{das} and prove that at the end of the $k+1$ round, agents
with all reliable interfaces reach a common decision. Thus by Lemma 7.5, if an
agent with all reliable interfaces receives a message by round $j$ $(j\leq k)$,
then by the end of round $j+1$, agents with all reliable interfaces reach a
common decision to attack. However on the other hand, by Lemma 7.6, if no agent
with all reliable interfaces receive any message by round $k$, then agents with
all reliable interfaces reach a common decision to retreat. Therefore, even if
the instigator has one or  more faulty interfaces, agents with all reliable
interfaces reach a common decision.
\end{enumerate}

\section{Results and Conclusion}
This investigation has lead us to the more formal design of secure communications
networks which are able to deal with both secure and insecure channels.
Maintaining the resiliency of the secure network is acheiveable given the use of a secret sharing
scheme, the ability to broadcast as well as using two-part secure lines. 
With our improvement to the treatment of $m/\phi$ and $\phi/m$ faults,
the security of a insecure communication channel can also be improved. Given
the $3m+1$ condition for insecure networks and the resilency improvement of
secure channels to $[(n-1)/2]$ a new design paradigm can be applied to networks
of agents. 
\\\\
Much work needs to be done on refining methods for obtaining 
Byzantine Agreements \cite{rush,lynch}. Further investigations into BA's could concentrate on:
\begin{itemize}
\item The correlation between the median voter theorem, a social choice
setting and BA's.
\item BA's and their use in Cluster networks.
\item BA's for establishing communication protocols with peers that are not always
available. (wandering solider problem \cite{lynch})
\end{itemize}


\begin{thebibliography}{199}

\bibitem{bar} Baranay, I. and Furedi, Z., \textit{Mental poker with three or more players} (1983), 
Information and Control, \textbf{59}, pp.84-93

\bibitem{ben} Ben-Or, M., Godwasser, S., Wigderson, A., \textit{Completeness
theorems for non-cryptographic fault-tolerant distributed computation} (1988),
Proceedings of the 6th Annual ACM Symposium on Theory of Computing, pp.379-386.

\bibitem{chor} Chor, B. and Kushilevitz, E. \textit{A zero-one law for Boolean
privacy}, (1991), SIAM Journal on Discrete Mathematics, vol.\textbf{4}, pp36-47

\bibitem{das} Dasgupta, P., \textit{Agreement under faulty interfaces} 
Information Processing Letters, \textbf{65} (1998) pp.125-129

\bibitem{dod} Dodis, Y., Halevi, S., and Rabin, T., \textit{A Cryptographic
Solution to a Game Theoretic Problem} Crypto 2000, Lecture Notes in Computer
Science, vol. \textbf{1880}, pp112-130, 2000

\bibitem{gold} Goldreich, O., Micali, S., Wigderson, A., \textit{How to play a 
mental game, or completeness theorem for protocols with honest} (1987) 
Proc. of the 19th ACM Symp. on Theory of Computing, pp.218-229

\bibitem{lam1} Lamport, L, Shostak, R., Pease, M., \textit{The Byzantine
Generals Problem} ACM Transactions on Progamming Languages and Syetems, vol.
\textbf{4} no. 3 pp382-401, July 1982 

\bibitem{lam2} Lamport, L, \textit{The Weak Byzantine Generals Problem}, 
(1983) Journal of the Association for Computing Macinery, 
vol.\textbf{30}, no. 3 pp668-676

\bibitem{lin} Linial, N., \textit{Chapter 38: Game-Theoretic Aspects of
Computing}, Handbook of Game Theory, With economic Applications vol. \textbf{2}
edited by Auman, R.J., Hart, S., Elsevier 1994

\bibitem{lynch} Lynch, N.A., \textit{Distributed Algorithms} (1996) Morgan
Kaufmann

\bibitem{men} Menezes, A., van Oorschot, P.C., Vanstone, S.A., \textit{Handbook of 
Applied Cryptography} CRC Press 1996

\bibitem{osb} Osbourne, M.J., Rubenstein, A., \textit{A Course in Game Theory}
MIT Press 1994

\bibitem{pease} Pease, M., Shostak, R., Lamport, L., \textit{Reaching Agreement
in the presence of Faults} Journal of the ACM, vol \textbf{27}, no. 2 pp.228-234 (1980)

\bibitem{ras} Rasmusen, E., \textit{Readings in Games and Information} (2001) 
Blackwell Publishers 

\bibitem{rush} Rushby, J., \textit{A Comparison of Bus Architectures for
Safety-Critical Embedded Systems} (2001) SRI International, CSL Technical Report
September 2001

\bibitem{schamir} Schamir, A., \textit{How to share a secret}, 
Communications of the ACM, vol. \textbf{22} pp.389-396 

\end{thebibliography}
\end{document}